\documentclass{article}
\usepackage{slashed,epsfig,amsmath,amssymb,enumitem}
\usepackage[papersize={8.5in,11in}]{geometry}
\renewcommand{\vec}[1]{\boldsymbol{\mathrm{#1}}}
\usepackage{graphicx}
\begin{document}
\title{Gravitational Renormalization Group Flow, Astrophysics and Cosmology}
\author{J. W. Moffat\\~\\
Perimeter Institute for Theoretical Physics, Waterloo, Ontario N2L 2Y5, Canada\\
and\\
Department of Physics and Astronomy, University of Waterloo, Waterloo,\\
Ontario N2L 3G1, Canada}
\maketitle

\begin{abstract}
A modified gravitational theory is developed in which the gravitational coupling constants $G$ and $Q$ and the effective mass $m_\phi$ of a repulsive vector field run with momentum scale $k$ or length scale $\ell =1/k$, according to a renormalization group flow. The theory can explain cosmological early universe data with a dark hidden photon and late time galaxy and cluster dynamics without dark matter. The theory agrees with solar system and binary pulsar observations.
\end{abstract}


\section{Introduction}

A modified gravity theory called Scalar-Tensor-Vector-Gravity {STVG}~\cite{Moffat} has been developed in which the gravitational coupling strength $G$, and the mass-range $\mu$ of a vector field $\phi_\mu$ were promoted to dynamical scalar fields in the action. The gravitational charge of the repulsive vector field is $Q=\sqrt{\alpha G_N}M$, where $\alpha$ is a parameter, $G_N$ is Newton's gravitational constant and $M$ is the mass. The structure growth and CMB data in early universe cosmology were explained by postulating a hidden, dark photon $\gamma'$ identified with the spin 1 vector field $\phi_\mu$ that plays the role of cold dark matter, non-thermally produced in the very early universe~\cite{Moffat2}. As the epoch of the formation of the first galaxies evolved the effective mass $\mu$ decreased in size and its late time value was fixed by fits to galaxy rotation curves and cluster dynamics without dark matter~\cite{MoffatRahvar1,TothMoffat,MoffatRahvar2} to be $\mu=0.042\,{\rm kpc}^{-1}$, corresponding to a mass $m_\phi=2.6\times 10^{-28}\,eV$. The bullet cluster dynamics has been explained without dark matter~\cite{BrownsteinMoffat}. By promoting $G$ and $\mu$ to be dynamical scalar fields in the action and field equations, it was necessary to postulate screening mechanisms to guarantee agreement with solar system and binary pulsar timing data~\cite{Moffat3}.

In the standard model of particle physics, the coupling constants $\alpha_i=g_i^2/4\pi$ run with momentum scale $k$ according to a renormalization group (RG) flow scenario based on a Wilsonian action. The coupling constants $\alpha_i(k)$ are {\it not promoted to be dynamical fields} in the Abelian and Yang-Mills actions. For the non-Abelian fields the screening leads to asymptotic freedom~\cite{Gross,Politzer}. In an earlier paper~\cite{Moffat4}, a Metric-Skewtensor-Tensor-Gravity (MSTG) theory was developed in which the effective coupling strengths $G$ and $\gamma_c$ of the gravitational field and a skew symmetric tensor field $F^{\mu\nu\lambda}$ were described by a a renormalization group flow formalism, analogous to the formalism in the standard model. A formulation of a gravity theory in which an effective average action $\Gamma_k$ is a ``course grained'' Wilsonian action function with a variable momentum scale $k$ has been developed by Reuter and collaborators~\cite{Reuter,Reuter2,Reuter3,Shapiro,Lauscher,Lauscher2,Saueressig}. The theory was formulated to yield asymptotic safe quantum gravity removing divergencies in the ultraviolet (UV)~\cite{Weinberg}.

In the following, we will adopt an effective running of the coupling constant $G(k)$, the gravitational charge $Q(k)$ and the mass $\mu(k)$ in a version of the modified gravity (MOG) theory, assuming that there is an underlying quantum theory of the classical action. Such a microscopic interpretation of the theory is not presently available, so we adopt a semi-phenomenological description of the scaling behavior of the MOG constants, which allows for an application of the model to the presently available early universe cosmology and late time astrophysical and astronomical data. This resolves the problem of the ``missing'' mass in the present universe, which is one of the most pressing problems in modern physics. As well as the metric $g_{\mu\nu}$, the repulsive gravitational vector field $\phi_\mu$ and its associated field strength $B^{\mu\nu}$ play a prominent role in the theory as dynamical fields. The gravitational coupling constants $G$ and $Q$, the cosmological constant $\Lambda$ and the vector field mass-range parameter $\mu$ are described as scale dependent running constants, obtained by solving appropriate RG flow equations.

In general relativity GR there is no known way to establish a Fourier transform linking momentum space to position space~\cite{Veltman,Ha,Moffat5}. A Fourier transform exists in fixed background symmetric spaces such as Minkowski spacetime and in Friedmann-Lema\^{i}tre-Robertson-Walker (FLRW) spacetime. The formal lack of a link between the particle physics momentum space and position space has been a source of conflict between GR and quantum gravity (QG) and this conflict can only be removed when a fully consistent QG is obtained. Making an analogy of the running of the gravitational coupling constants such as $G(k)$ with quantum field theory (QFT) and quantum chromodynamics (QCD) only applies in a semi-phenomenological framework. We describe the effective action $\Gamma_k$ as a course-grained average description in the RG flow formalism. The effective course-grained approach advocated by Reuter and collaborators can provide a phenomenological RG flow treatment of gravity that can be expected to be valid in nature.

\section{\bf The Gravitational Action and Field Equations}

The diffeomorphism invariant action for the theory is given by
\begin{equation}
S=S_G+S_\phi+S_{\phi M}+S_M,
\end{equation}
where $S_G$ is the Einstein-Hilbert action (we choose units such that $c=1$):
\begin{equation}
\label{gravaction} S_G=\frac{1}{16\pi G}\int d^4x\sqrt{-g}(R+2\Lambda),
\end{equation}
where $g={\rm det}(g_{\mu\nu})$, $g_{\mu\nu}$ is the symmetric
metric tensor of pseudo-Riemannian geometry, $R=g^{\mu\nu}R_{\mu\nu}$ is the Ricci scalar and $\Lambda$ is the
cosmological constant. Moreover, $S_\phi$ is the field action:
\begin{equation}
\label{phiaction} S_\phi=-\frac{1}{4\pi}\int d^4x\sqrt{-g}\biggl(\frac{1}{4}B^{\mu\nu}B_{\mu\nu}+V(\phi_\mu)\biggr),
\end{equation}
where
\begin{equation}
B_{\mu\nu}=\partial_\mu\phi_\nu-\partial_\nu\phi_\mu.
\end{equation}
We shall choose the potential $V(\phi_\mu)$ to be
\begin{equation}
V(\phi_\mu)=-\frac{1}{2}\mu^2\phi^\mu\phi_\mu +W(\phi_\mu),
\end{equation}
where $W(\phi_\mu)$ denotes a $\phi_\mu$ field self-interaction potential. We can choose as a model of the self-interaction potential:
\begin{equation}
W(\phi_\mu)=\frac{1}{4}h(\phi^\mu\phi_\mu)^2,
\end{equation}
where $h$ is a coupling constant.

The gravitational constant $G$ in the action $S_G$ is defined in terms of the ``bare'' Newtonian constant $G_N$:
\begin{equation}
\label{gravirenorm} G=G_NZ,
\end{equation}
where $Z=1+\alpha$ corresponds to a ``renormalization'' of $G$ and $\alpha$ is a dimensionless parameter.

The actions $S_\phi$, $S_{\phi M}$ and $S_M$ satisfy the relations
\begin{equation}
\frac{1}{\sqrt{-g}}\frac{\delta S_M}{\delta
g^{\mu\nu}}=-\frac{1}{2}T_{M\mu\nu},\quad\frac{1}{\sqrt{-g}}\frac{\delta
S_\phi}{\delta g^{\mu\nu}}=-\frac{1}{2}T_{\phi\mu\nu},\quad\frac{\delta S_{\phi M}}{\delta\phi_\mu}=-J^\mu.
\end{equation}
Here, $T_{M\mu\nu}$ is the energy-momentum tensor for matter, while $T_{\phi\mu\nu}$ is the energy-momentum tensor containing the
contributions from the massive vector field $\phi_\mu$:
\begin{equation}
\label{Tphi}
T_{\phi\mu\nu}=\frac{1}{4\pi}\biggl[{B_\mu}^\alpha B_{\nu\alpha}+2\frac{\partial V(\phi_\mu)}{\partial g^{\mu\nu}}-g_{\mu\nu}\biggl(\frac{1}{4}B^{\alpha\beta}B_{\alpha\beta}+V(\phi_\mu)\biggr)\biggr].
\end{equation}
Moreover, $J^\mu=(J^0,J^i)\,(i=1,2,3)$ is the current density for the $\phi_\mu$ field gravitational charge:
\begin{equation}
J^\mu=\sqrt{\alpha G_N}\rho_M u^\mu,
\end{equation}
where $\rho_M$ is the density of matter and $u^\mu=dx^\mu/ds$. We have
\begin{equation}
\label{Qequation}
Q=\sqrt{\alpha G_N}M = \int d^3x J^0.
\end{equation}

The field equations derived from the action principle are given by
\begin{equation}
\label{Gequation} G_{\mu\nu}-\Lambda g_{\mu\nu}=8\pi GT_{\mu\nu},
\end{equation}
\begin{equation}
\label{phifieldequations}
\nabla_\nu B^{\mu\nu}+\frac{\partial V(\phi_\mu)}{\partial\phi_\mu}=-J^{\mu},
\end{equation}
and
\begin{equation}
\label{Bcurleq}
\nabla_\sigma B_{\mu\nu}+\nabla_\mu B_{\nu\sigma}+\nabla_\nu B_{\sigma\mu}=0.
\end{equation}
Here, $G_{\mu\nu}=R_{\mu\nu}-\frac{1}{2}g_{\mu\nu}R$, $T_{\mu\nu}=T_{M\mu\nu}+T_{\phi\mu\nu}$ and $\nabla_\sigma$ denotes
the covariant derivative with respect to $g_{\mu\nu}$. The Bianchi identities satisfied by the Einstein tensor, $\nabla_\nu G^{\mu\nu}=0$, lead to the conservation laws:
\begin{equation}
\nabla_\nu T^{\mu\nu}=0.
\end{equation}

\section{Renormalization Group Flow and Running Coupling and Mass Constants}

From the point of view of QFT, we know that the electric charge $e$ or the fine structure constant $\alpha_{QED}$ are not really constant but they are scale dependent or ``running'' quantities. In quantum electrodynamics (QED) the coupling constant $\alpha_{QED}$ depends on the renormalization scale $k$, a parameter with the dimension of momentum. The QED vacuum is a sea of virtual positron-electron pairs. A test charge is screened by a cloud of positron-electron pairs surrounding it, so that at smaller distances the test charge appears to be larger due to screening and smaller at large distances. In the non-Abelian Yang-Mills theory or quantum chromodynamics (QCD), the strong coupling constant $\alpha_s=\alpha_s(k)$ is anti-screened, so that at large distances $\alpha_s(k)$ grows with distance. In analogy with the situation in QCD, we can envisage that the gravitational coupling constant $G=G(k)$ is also anti-screened, because of the non-Abelian nature of the gravitational field. Then, $G=G(k)$ is expected to grow in size as the distance scale increases due to non-perturbative renormalization effects.

We have postulated a renormalized Newtonian gravitational constant $G=G_NZ$ in our action Eq.(\ref{gravaction}):
\begin{equation}
\label{Grenormalized}
G\equiv GZ=G_N(1+\alpha).
\end{equation}
We shall now implement the running of the effective $G$ and the $B_{\mu\nu}$ field gravitational coupling constant $\alpha_Q=Q^2/4\pi$ (${\hbar}=1$ and $Q=\sqrt{\alpha G_N}M$) by using RG flow arguments in which the ``classical'' coupling constants $G$ and $Q$ possess a scale dependent running behavior obtained by solving appropriate RG flow equations.

We postulate scale dependent effective actions $\Gamma_k[g_{\mu\nu}]$ and $\Gamma_k[\phi_\mu]$, corresponding
to ``coarse-grained'' free energy functionals, which define an effective field theory valid at the scale $k$ or length scale
$\ell=1/k$, where $\ell$ is the length scale resolution of the observed system. Then, the $\Gamma_k$ are the bare actions obtained by integrating out all quantum fluctuations with momenta larger than an infrared (IR) cutoff $k_{IR}$, or wavelengths smaller than $\ell$. The $\Gamma_k$ are determined by functional differential equations for the RG flow, including the dimensionless coupling constants ${\bar G}(k)=k^2G(k), \alpha_Q(k)$, the cosmological constant $\lambda(k)=\Lambda(k)/k^2$ and the mass ${\bar\mu}=\mu(k)/k$. The dimensionless constant $G(k)\Lambda(k)={\bar G}(k)\lambda(k)$ is the quantity that is determined by measurement. We observe from (\ref{Qequation}) and (\ref{Grenormalized}) that both $G(k)$ and $\alpha_Q(k)$ are determined by the running of $\alpha(k)$. The RG flow equations for $\alpha(k)$, $\lambda(k)$ and $\bar\mu(k)$ are described by the ordinary coupled differential equations:
\begin{equation}
k\frac{d}{dk}\alpha(k)=\beta_\alpha(\alpha,\lambda,{\bar\mu}),\quad k\frac{d}{dk}\lambda(k)=\beta_\lambda(\alpha,\lambda,{\bar\mu}),\quad
k\frac{d}{dk}\bar\mu(k)=\beta_{\bar\mu}(\alpha,\lambda,{\bar\mu}).
\end{equation}
To solve the problem, we are required to restrict the RG flow to a finite-dimensional subspace~\cite{Reuter,Saueressig}, corresponding to a
truncated theory space. In the truncated space only the coupling constants ${\bar G}$ and $\alpha_Q$, the cosmological constant
$\lambda$ and the mass $\bar\mu$ are considered.

Every solution ${\bar G}(k),\alpha_Q(k),\lambda(k),{\bar\mu(k)}$ of the truncated flow equations is associated with the
family of action functionals:
\begin{equation}
\Gamma_k[g_{\mu\nu},\phi_\mu]=\Gamma_k[g_{\mu\nu}]+\Gamma_k[\phi_\mu].
\end{equation}
The effective average action formalism follows the RG flow from the bare action $\Gamma_{k=\infty}=S$, corresponding to the
initial condition for the $\Gamma_k$-flow equation, down to $k=0$ and $\Gamma_{k=0}=\Gamma$ corresponding to the effective action.
We have to solve the effective equations of motion
\begin{equation}
\frac{\delta\Gamma_k[g_{\mu\nu}]}{\delta g_{\mu\nu}}=0,\quad \frac{\delta\Gamma_k[\phi_\mu]}{\delta\phi^\mu}=0.
\end{equation}
In practice, we truncate the space of solutions and for the UV sector, simple local truncations are sufficient to describe the
physical system, but in the IR limit $k\rightarrow 0$ {\it nonlocal} terms must be included in the truncation procedure.

We shall presently just consider the running of the coupling constants $G(k)$ and $\alpha_Q(k)$. The RG flow equations are
dominated by two fixed points ${\bar G}_*$ and $\alpha_{Q*}$: Gaussian fixed points at ${\bar G}_*=\alpha_{Q*}=0$ and
non-Gaussian ones with ${\bar G}_* > 0$ and $\alpha_{Q*} > 0$. The high-energy, short distance
behavior is governed by the non-Gaussian fixed points, so that for $k\rightarrow\infty$ all
the RG trajectories run into these fixed points by taking the UV cutoff along a trajectory running into the fixed points. This could lead to a non-perturbatively renormalizable quantum gravity theory. The conjectured existence of such a non-perturbatively renormalizable quantum gravity theory still has to be demonstrated.

The running of the constants $G(k)$ and $\alpha_Q$ is described by
\begin{equation}
k\partial_k G(k)=\eta_GG(\alpha(k)),\quad k\partial_k\alpha_Q(k)=\eta_Q\alpha_Q(\alpha(k)).
\end{equation}
If $\eta_G > 0$ and $\eta_Q > 0$, then $G(k)$ and $\alpha_Q(k)$ increase with increasing mass scale $k$ (or decreasing length scale $\ell=k^{-1}$ as $k\rightarrow\infty$), while they decrease for $\eta_G < 0$ and $\eta_Q < 0$. In analogy with gauge theory, the first case refers to ``screening'', while the second refers to ``anti-screening''. Both the $G(k)=G_N(1+\alpha(k))$ and $\alpha_Q(k)=\alpha(k)G_NM^2/4\pi$ are determined by the one free parameter $\alpha(k)$. We adopt the ans{\"a}tz that $\alpha(\ell=k^{-1})$ has the behavior shown in Fig. 1. It is small at the scale of the solar system, so that $G\sim G_N$ and $Q\sim 0$ which guarantees agreement with all the solar system experiments. It also guarantees agreement with the timing data for binary pulsars.

As in QCD, we see that if we interpret $\ell=1/k$ as a distance scale, then both $G(k)$ and $\alpha_Q(k)$ increase all the way from the UV energy scale to the IR energy scale as $k\rightarrow 0$. This corresponds to a quantum gravity {\it anti-screening}, playing the analogous role to the anti-screening in Yang-Mills QCD. This interpolates between the constant Newtonian value $G(k_{\rm lab})\sim G_N$ at large $k$ corresponding to short distances and the increasingly larger values of G(k) at larger distances for small $k$. We can as in QCD eliminate an explicit dependence of $G(k)$ and $\alpha_Q(k)$ on a cutoff $k_C$ by choosing a renormalization or reference momentum $m_R$. The equations for $G(k)$ and $\alpha_Q(k)$ now contain only finite, physically measurable quantities depending on $k$ and $m_R$. In our formulation of the RG flow, we do not necessarily assume that the asymptotic freedom condition $G(k)\rightarrow 0$ as $k\rightarrow\infty$ and that a non-Gaussian fixed point exists.

\begin{figure}
\centering \includegraphics[scale=0.3]{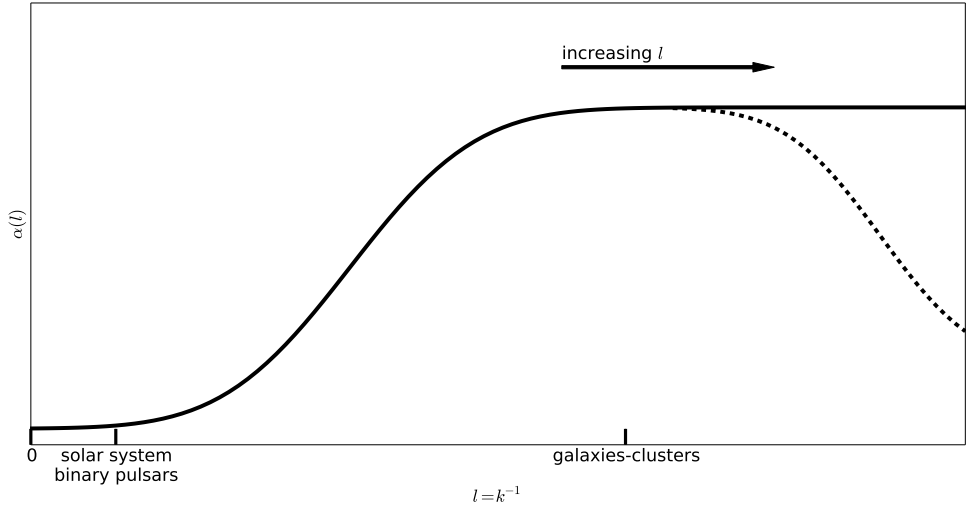}\\
\caption{\label{fig:alpha_v_l}A schematic depiction of the running of $\alpha(\ell)$ $(\ell=k^{-1})$. The dashed curve shows the behavior of $\alpha(\ell)$ for $\alpha(\ell)\rightarrow 0$ as $\ell\rightarrow\infty.$}
\end{figure}

\section{Galaxy and Cluster Dynamics}

We expand $g_{\mu\nu}$ around the Minkowski spacetime metric $\eta_{\mu\nu}$:
\begin{equation}
g_{\mu\nu}=\eta_{\mu\nu}+h_{\mu\nu},
\end{equation}
where $|h_{\mu\nu}| << 1$. The action for a test particle is
\begin{equation}
S_{\rm tp}=-\biggl( m\int d\tau+q\int\phi_\mu \frac{dx^\mu}{d\tau}\biggr),
\end{equation}
where $m$ and $q=\sqrt{\alpha G_N}m$ are the test particle mass and gravitational charge, respectively, and $\phi_\mu=(\phi_0,\phi_i)$.
The spherically symmetric static solution for $\phi_0(r)$ is obtained from the equation ($\phi_0'=d\phi_0/dr$):
\begin{equation}
\phi_0''+\frac{2}{r}\phi_0'-\mu^2\phi_0=0.
\end{equation}
The solution is given by
\begin{equation}
\label{phisolution}
\phi_0(r)=-Q\frac{\exp(-\mu r)}{r},
\end{equation}
where the charge $Q=\sqrt{\alpha G_N}M$.

The equation of motion of a test particle is given by
\begin{equation}
\frac{d^2x^\mu}{d\tau^2}+{\Gamma^\mu}_{\alpha\beta}\frac{dx^\alpha}{d\tau}\frac{dx^\beta}{d\tau}=\frac{q}{m}{B^\mu}_\nu\frac{dx^\nu}{d\tau},
\end{equation}
where $\tau$ is the proper time along the particle trajectory. We assume that in the slow motion approximation $dr/d\tau\sim dr/dt$ and $2GM/r < 1$, then for the radial acceleration of the test particle we get
\begin{equation}
\frac{d^2r}{dt^2}+\frac{GM}{r^2}=\frac{qQ}{m}\frac{\exp(-\mu r)}{r^2}(1+\mu r).
\end{equation}
For $ qQ/m=\alpha G_NM$ and $G=G_N(1+\alpha)$, the modified Newtonian acceleration law for a point particle is given by~\cite{Moffat,MoffatRahvar1}:
\begin{equation}
\label{accelerationlaw}
a(r)=-\frac{G_NM}{r^2}\biggl\{1+\alpha\biggl[1-\exp(-r/r_0)\biggl(1+\frac{r}{r_0}\biggr)\biggr]\biggr\},
\end{equation}
where $r_0=1/\mu$.  The acceleration law can be extended to a distribution of matter:
\begin{equation}
\label{accelerationlaw2}
a({\vec x})=-G_N\int d^3{\vec x}'\frac{\rho({\vec x}')({\vec x}-{\vec x}')}{|{\vec x}-{\vec x}'|^3}
[1+\alpha-\alpha\exp(-\mu|{\vec x}-{\vec x}'|)(1+\mu|{\vec x}-{\vec x}'|)].
\end{equation}
The parameter $\alpha$ will run with distance scale in the RG flow scenario as shown in Fig. 1.
At each distance scale $k=1/\ell$ the parameter $\alpha(k=1/\ell)$ is treated as a constant, so that at a given physical scale $\ell_{\rm phys}$ the acceleration law (\ref{accelerationlaw}) and (\ref{accelerationlaw2}) is valid. At the distance scale corresponding to the solar system, $\ell_{\rm solar}\sim 40\, AU$, $\alpha(\ell) << 1$, so that solar system experiments including the bending of light by the sun and the Cassini probe time delay observation agree with GR. Moreover, the binary pulsar timing observations, for distance scales corresponding to the compact pulsar dimensions and $\ell\leq 8\,{\rm kpc}$, will agree with the observations of binary pulsar timings for $\alpha(\ell) << 1$. For galactic and cluster scales $\ell_G$, fits to galaxy rotation curves and cluster dynamics are obtained with $\alpha=8.89\pm 0.34$ and $\mu=0.042\pm 0.004\,{\rm kpc}^{-1}$ without dark matter~\cite{MoffatRahvar1,TothMoffat,MoffatRahvar2}. Moreover, the bullet cluster data can be fitted without dark matter~\cite{BrownsteinMoffat}.

\section{Cosmology}

A description of early universe cosmology is obtained by adopting the homogeneous and isotropic FLRW background metric:
\begin{equation}
ds^2=dt^2-a^2(t)\biggl(\frac{dr^2}{1-Kr^2}+r^2(d\theta^2+\sin^2\theta d\phi^2)\biggr),
\end{equation}
where $K=-1,0,1\,({\rm length}^{-2})$ for open, flat and closed universes, respectively. We use the energy-momentum tensor of a perfect fluid:
\begin{equation}
T_{\mu\nu}=(\rho+p)u_\mu u_\nu-pg_{\mu\nu},
\end{equation}
where $T_{\mu\nu}$ is the total energy-momentum tensor in our RG flow framework, and $\rho$ and $p$ are the density and pressure, respectively. In the following, we assume a spatially flat universe $K=0$. In the background FLRW spacetime, we have $\phi_0\neq 0, \phi_i=0$ and $B_{\mu\nu}=0$. Moreover, we have
\begin{equation}
\rho=\rho_M+\rho_\phi,\quad p=p_M+p_\phi,
\end{equation}
where $\rho_i$ and $p_i$ denote the components of density and pressure associated with the matter and the $\phi_\mu$ vector field.

The Friedmann equations are given by
\begin{equation}
\label{Friedmann3}
H^2=\frac{8\pi G\rho}{3}+\frac{\Lambda}{3},
\end{equation}
\begin{equation}
\frac{\ddot a}{a}=-\frac{4\pi G}{3}(\rho+3p)+\frac{\Lambda}{3},
\end{equation}
where $H=\dot a/a$. The energy conservation equation is
\begin{equation}
\dot\rho+3\frac{d\ln a}{dt}(\rho+p)=0.
\end{equation}

A linear perturbation on the FLRW background will connect the theory with anisotropy observations in the CMB. In the conformal metric with the time $d\eta=dt/a(t)$:
\begin{equation}
ds^2=a^2(\eta)(d\eta^2-d{\vec x}^2),
\end{equation}
the metric perturbations are defined in the Newtonian gauge:
\begin{equation}
g_{00}({\vec x},t)=a^2(t)(1+2\Phi({\vec x},t)),\quad g_{ij}({\vec x,t})=a^2(t)(1-2\Phi({\vec x},t)\delta_{ij},
\end{equation}
where $\Phi$ denotes the gravitational potential. The perturbations of the vector field $\phi_\mu$ are defined by
\begin{equation}
\phi_\mu({\vec x},t)=a(t)(\bar\phi_\mu(t)+\delta\phi_\mu({\vec x},t))\,(\bar\phi_i(t)=0),
\end{equation}

In a purely baryonic early universe, the coupling of baryons to photons before and during the recombination era will suffer damping, causing the collisional propagation of radiation from overdense to underdense regions. In order to obtain the expansion of the universe and structure growth to agree with observation, we invoke a cold dark matter in the form of the neutral field $\phi_\mu$, which plays the role of a ''dark photon'' $\gamma'$ with a sufficiently big mass $m_\phi$. The dark photon $\gamma'$ interacts weakly with standard model particles through its gravitational strength coupling $Q=\sqrt{\alpha G_N}m_\phi$ and it has zero pressure. In our cosmology model we have
\begin{equation}
\Omega_b=\frac{8\pi G_N\rho_b}{3H^2},\quad \Omega_\phi=\frac{8\pi G_N\rho_\phi}{3H^2},\quad \Omega_\Lambda=\frac{8\pi G_N\rho_{\rm vac}}{3H^2},
\end{equation}
where $\rho_b$ and $\rho_{\rm vac}$ denote the density of baryons and vacuum density, respectively. We also have a contribution from neutrinos:
\begin{equation}
\Omega_\nu=\frac{8\pi G_N\rho_\nu}{3H^2},
\end{equation}
where $\rho_\nu$ denotes the density of neutrinos. We assume that the dark photon density $\rho_\phi$ dominates before the formation of the first galaxies, and the acceleration of the universe expansion is determined by the cosmological constant $\Lambda$.

In Figure 2, we depict the running of $\alpha(a(t))$ as the cosmic scale factor $a(t)$ increases with time. We choose $k=\xi/a(t)$ where $\xi$ is a constant and before the formation of the first galaxies, $\rho_\phi > \rho_b$. We have for $G=G_N(1+\alpha)$ that $G\sim G_N$ and $\alpha < 1$ before galaxy formation. After the epoch of galaxy formation $\rho_b > \rho_\phi$ and the {\it baryon density begins to dominate the density of matter}. In Figure 3, we schematically depict the running of $m_\phi$ with increasing $a(t)$, when it reaches the value $m_\phi=2.6\times 10^{-28}\,eV$ corresponding to the value $\mu=0.042\,{\rm kpc}^{-1}$ or $1/\mu=24\,{\rm kpc}$ used to fit galaxy rotation curves and the dynamics of galaxy clusters.

The Newtonian potential in the early universe is given by
\begin{equation}
k^2\Phi=4\pi G_N(\rho_b\delta_b+\rho_{\rm CDM}\delta_{\rm CDM}),
\end{equation}
where $k^2$ denotes the square of the wave number, $\rho_{\rm CDM}=\rho_\phi$ and $\delta_i$ denotes the perturbation density contrast for the baryon and CDM components. If $\rho_\phi$ is sufficiently large, then $\delta_\phi=\delta_{\rm CDM}$ will not be erased and the cosmological model yields agreement with data for the structure growth, rate of expansion of the universe, the CMB angular acoustical power spectrum. We show in Figure 4, the best fit MOG ($\Lambda$ CDM model) TT power spectrum~\cite{Planck}. The formation of galaxies about 4-5 million years after the big bang occurs during the time when the modified gravity theory takes over with $\rho_b > \rho_\phi$ and galaxy dark matter haloes do not form. The data for the matter power spectrum determined by galaxy correlation functions can be fitted within modified gravity without dark matter~\cite{MoffatToth,Moffat2}.\\

\begin{figure}
\centering \includegraphics[scale=0.3]{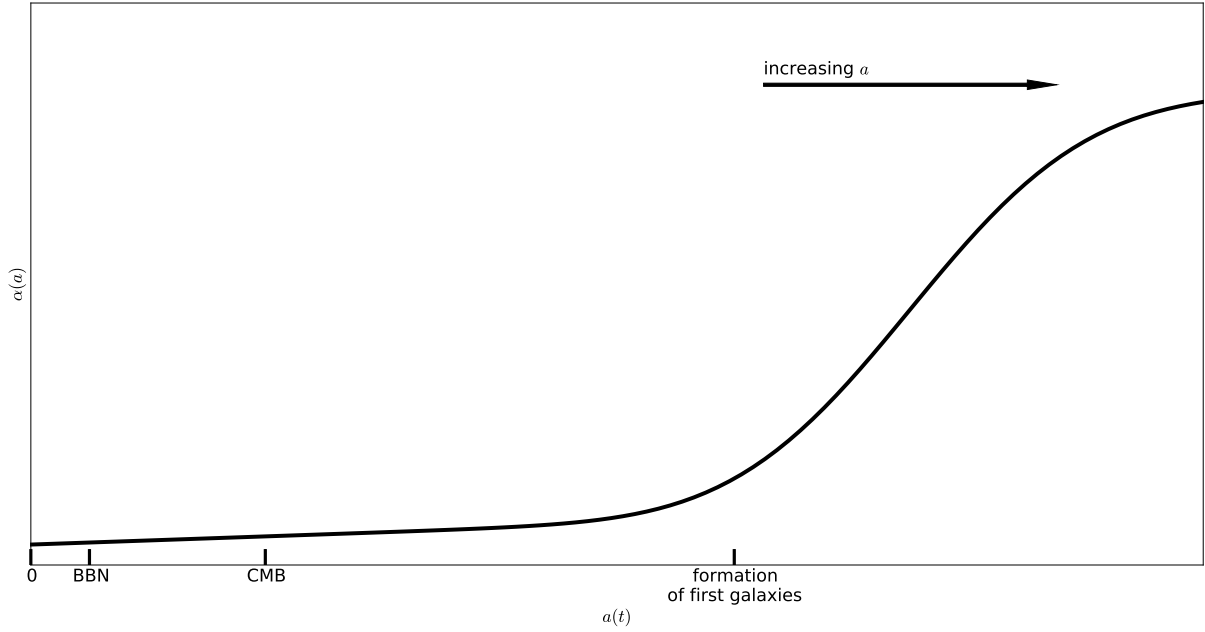}\\
\caption{\label{fig:alpha_v_a}Schematic running of $\alpha=\alpha(a(t))$, where $a(t)$ denotes the cosmological scale factor. As $a(t)\rightarrow 0$ $\alpha\rightarrow 0$ according to the identification $k=\xi/a(t)$ and $G\rightarrow G_N$ in the Cosmic Microwave Background (CMB) and Big Bang Nucleosynthesis (BBN) epochs.}
\end{figure}

\begin{figure}
\centering \includegraphics[scale=0.3]{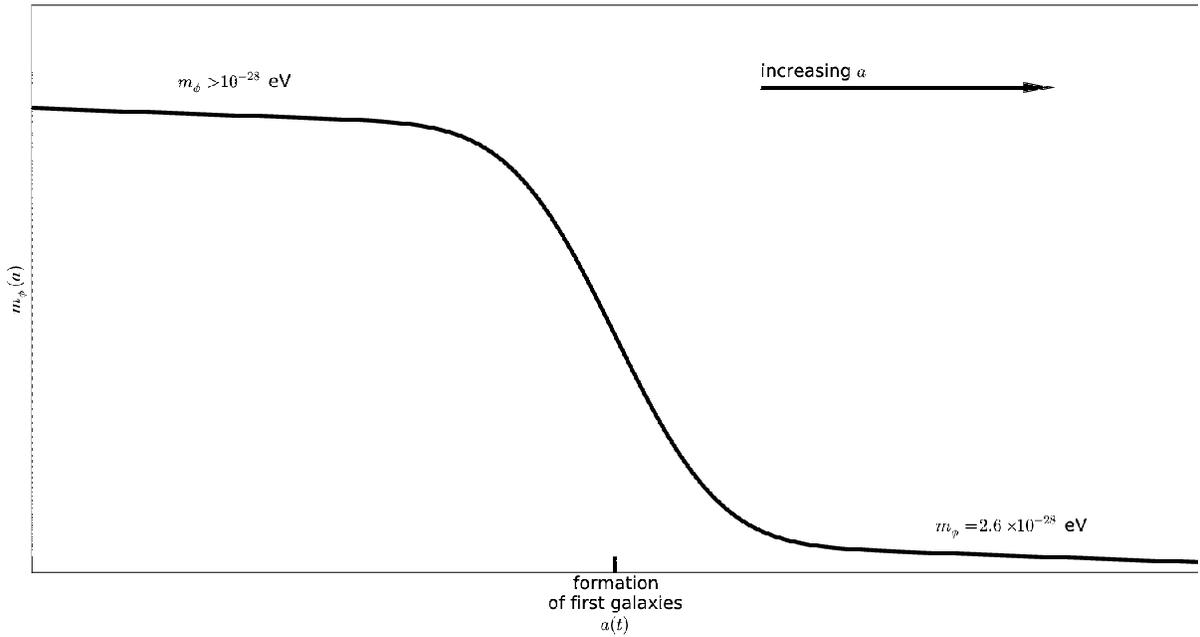}\\
\caption{\label{fig:scalar-field-mass}Schematic running of the vector field mass $m_\phi(a(t))$ as $a(t)\rightarrow\infty$.}
\end{figure}

\begin{figure}
\centering \includegraphics[scale=0.75]{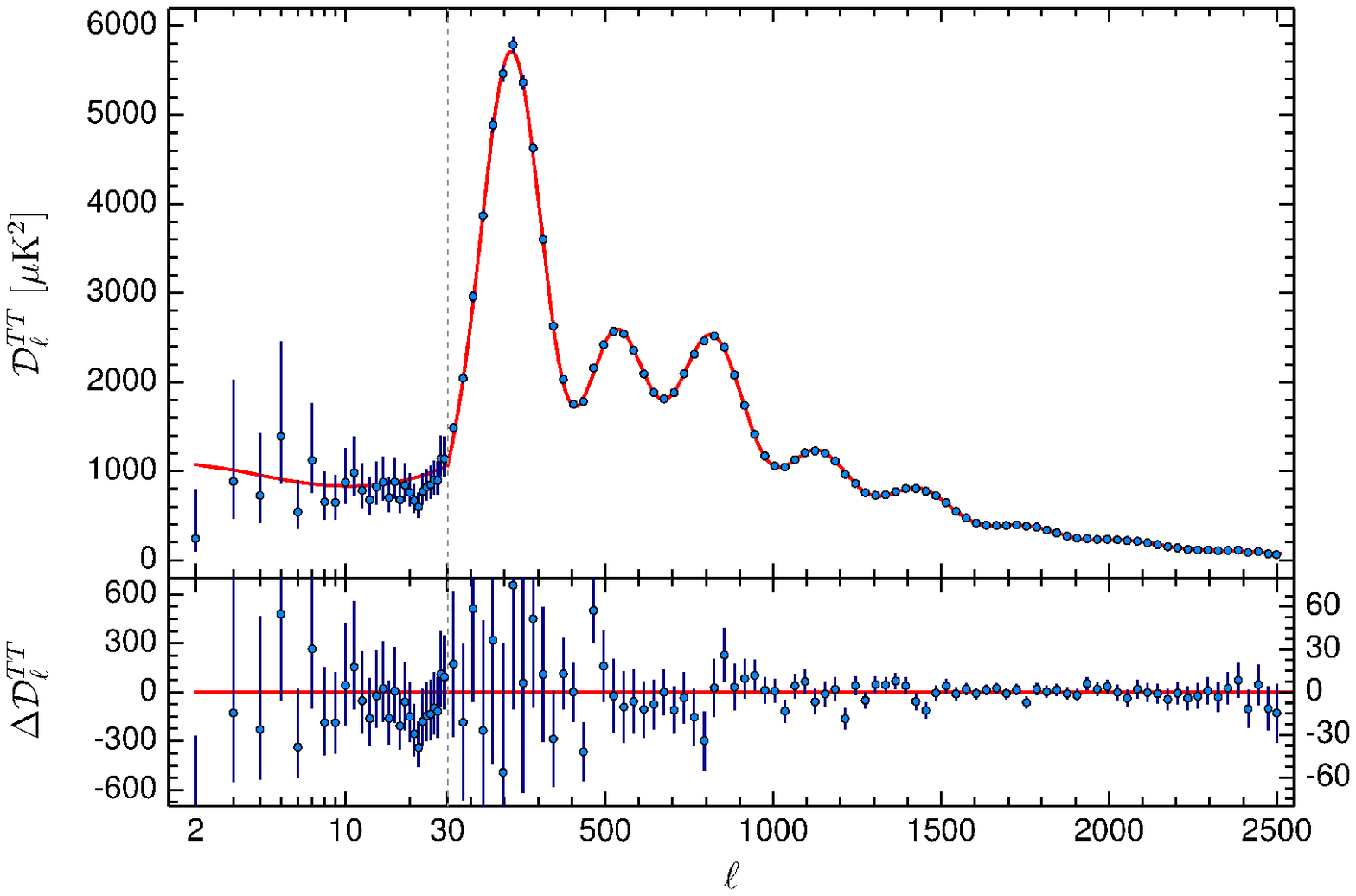}\\
\caption{\label{fig:angularpowerspectrum}The TT power spectrum using $\Omega_bh^2, \Omega_c h^2=\Omega_\phi h^2,\Omega_\Lambda$ and the remaining parameters of the best fit listed in Table 4 of the Planck 2015 paper~\cite{Planck}. The red line shows the best-fit MOG spectrum (and the $\Lambda$CDM model spectrum). The lower panel shows the residuals with respect to the MOG model.}
\end{figure}

We have revived the original proposal for modified gravity that $G$, $\alpha_Q$ and $\mu$ be considered as running parameters, as in
RG flow~\cite{Moffat4}. With this interpretation, the applicable values of the parameters in a particular situation depend on how the
observed system is probed by telescopes and satellites, and the corresponding spatial resolution. Observations of a distant galaxy require a resolution of order kpcs, and the parameter values $\alpha$ and $\mu$ will correspond to that scale. If one observes objects on Earth or in the solar system, then the spatial resolution will be of order AU or less.  This is analogous to measurements of running coupling constants in particle physics, where
the scale is the inverse of the momentum transfer, and higher energy collisions probe smaller scales. The geometry determined by the metric and curvature of spacetime is resolved spatially by observation and objects farther from observing instruments are not as well resolved as closer objects such as our solar system.

\section{\bf Conclusions}

We have developed a gravity theory based on a modification of GR in which the action has in addition to the Einstein-Hilbert action for the metric $g_{\mu\nu}$ a dynamical action for a massive vector field $\phi_\mu$ and the field strength $B_{\mu\nu}$. The gravitational coupling strengths $G(k)$ and $\alpha_Q(k)$ run with momentum $k$ in analogy with the running coupling constants in Yang-Mills theory and QCD. In our applications to astrophysics and cosmology, the $G(k)$ and $\alpha_Q(k)$ are anti-screened in the physical space of $G, \alpha_Q$ and $\Lambda$. The running of the coupling constants $G(k)$ and $\alpha_Q(k)$ is determined by Wilsonian RG trajectories. The trajectories have Gaussian fixed points when the $\beta_G$ and $\beta_{\alpha_Q}$ functions vanish at the points $G_*$ and $\alpha_{Q*}$. Non-Gaussian fixed points can occur for non-zero values of $G_*$ and $\alpha_{Q*}$. If they occur on the RG flow trajectories, it can be argued that unphysical divergences do not occur for UV values of $k$~\cite{Reuter,Reuter2,Saueressig}. The running of $G(\ell)$ and of $\alpha_Q(\ell)$ is determined by the running of the dimensionless constant $\alpha(\ell)$, such that for small distance scales corresponding to the solar system or binary pulsars $\alpha << 1$. For cosmology the running of $\alpha$ is controlled by the magnitude of the cosmic scale $a(t)$ according to the scaling parameter $k=\xi/a(t)$. In the early universe before the formation of the first galaxies $\alpha < 1$, and the mass $m_\phi$ of the vector field $\phi_\mu$ is big enough to allow the vector particle be a hidden dark matter photon $\gamma'$, while for the late-time universe $m_\phi\sim 10^{-28}\,eV$. In the late-time epoch the density $\rho_\phi$ is diluted and $\rho_\phi < \rho_b$, so that the baryon density $\rho_b$ dominates and the dark photon $\gamma'$, coupled with gravitational strength to standard model particles, is {\it undetectable}. The rotation curves of galaxies and the cluster dynamics, including cluster collisions such as the Bullet Cluster, are fitted without dark matter~\cite{MoffatRahvar1,TothMoffat,MoffatRahvar2,BrownsteinMoffat}.

The RG flow gravity theory and the determination of the running of the parameter $\alpha$ are expected to be reliable for {\it weak gravitational fields}, while for the strong gravitational fields of black holes in the present universe~\cite{Moffat6,Moffat7,Moffat8} the running value of $\alpha$ will require a non-perturbative treatment outside the scope of the presently developed theory.

For appropriate running of the constants $G(k)$, $\alpha_Q$ and $m_\phi$ for physical values of the scaling parameter $k$ or length scale $\ell=k^{-1}$, observational data for the solar system, binary pulsars, galaxies, galaxy clusters and cosmology can be fitted by the theory.

\vskip 0.2
true in {\bf Acknowledgments}
\vskip 0.2 true in

I thank Martin Green and Viktor Toth for helpful discussions and Jonah Miller for preparing the graphics. Research at the Perimeter Institute for Theoretical Physics is supported by the Government of Canada through industry Canada and by the Province of Ontario through the Ministry of Research and Innovation (MRI).

\vskip 0.5 true in

\end{document}